\documentclass{ws-rv9x6}

\usepackage{subfigure,ws-rv-van}

\begin{document}

\chapter{Partial Conservation Law in a Schematic Single $j$ Shell
  Model}

\author{Wesley Pereira, Ricardo Garcia, Larry Zamick and Alberto
  Escuderos}

\address{Department of Physics and Astronomy, Rutgers University, \\
  Piscataway, New Jersey 08854, USA}

\author[W. Pereira, R. Garcia, L. Zamick, A. Escuderos and
  K. Neerg\aa rd]{Kai Neerg\aa rd}

\address{Fjordtoften 17, 4700 N\ae stved, Denmark}

\begin{abstract}
  We report the discovery of a partial conservation law obeyed by a
  schematic Hamiltonian of two protons and two neutrons in a $j$
  shell. In our Hamiltonian the interaction matrix element of two
  nucleons with combined angular momentum $J$ is linear in $J$ for
  even $J$ and constant for odd $J$. It turns out that in some
  stationary states the sum of the angular momenta $J_p$ and $J_n$ of
  the proton and neutron pairs is conserved. The energies of these
  states are given by a linear function of $J_p + J_n$. The
  systematics of their occurrence is described and explained.
\end{abstract}

\body

\section{Introduction}

Among the many contributions of Gerry Brown to Nuclear Physics one of
the first that comes to the minds of many is his development with Tom
Kuo of realistic nuclear matrix elements.\cite{ref:KuBr} These involve
the very complicated nucleon nucleon interaction and the added
complication of handling the hard core by obtaining a $G$ matrix which
a researcher could easily handle. However our present work is inspired
by another aspect of Gerry Brown's contributions---his use of simple
schematic models to bring out the physics of the more complex
calculations. One example is his early article with Marc Bolsterli in
Physical Review Letter on dipole states in nuclei.\cite{ref:BrBo}
Their simple model employs a delta interaction with radial integrals
set to a constant. One state gets elevated to a high energy and
contains all the dipole strength. Gerry and Marc compared their
results with a more detailed calculation of Elliott and
Flowers.\cite{ref:ElFl} These authors obtained two collective states,
and Gerry and Marc noted that a defect of their model was the neglect
of the spin orbit interaction. However they expected that it could
work better for heavier nuclei. A quote from the end of their paper:
``The schematic model is of course no substitute for detailed
calculations but indicates the possibility of these coherent features
in a simple way.''

In Gerry's first book \textit{Unified Theory of Nuclear
  Models}\cite{ref:UnTh} he discusses besides more elaborate schemes
of calculation such schematic models as Elliott's SU(3)
model to describe nuclear rotation\cite{ref:El} and Racah's seniority
scheme displaying the physics of pairing in nuclei.\cite{ref:Ra}

Below we consider a simple model with only one $j$ shell, where we put
both protons and neutrons. Such a model was applied in the early days
to the description of nuclear spectra, magnetic moments, beta deay
\textit{etc.} in the $1f_{7/2}$
shell.\cite{ref:BaCuZa,ref:CuBaZa,ref:GiFr,ref:Gi1,ref:Gi2} The
interaction matrix elements were taken from the spectra of $^{42}$Ca
and $^{42}$Sc. The $^{42}$Sc, $T = 0$ spectrum was poorly known at
that time and some of the assignments were wrong. Revised matrix
elements were later extracted from the correct $^{42}$Sc spectrum by
Zamick and Robinson,\cite{ref:ZaRo} and these matrix elements were
employed by Escuderos, Zamick and Bayman in complete
calculations for the $1f_{7/2}$ shell.\cite{ref:EsZaBa} Despite large
differences between the original and revised matrix elements,
especially a lowering of those for two nucleon angular momentum %
$J = 1$, 3 and 5 by about half an MeV, no red flags were raised. This
indicates a certain insensitivity to the $T = 0$ matrix, a theme that
will pervade this work.

\begin{figure}\label{fig:42Sc}
  \centerline{\includegraphics{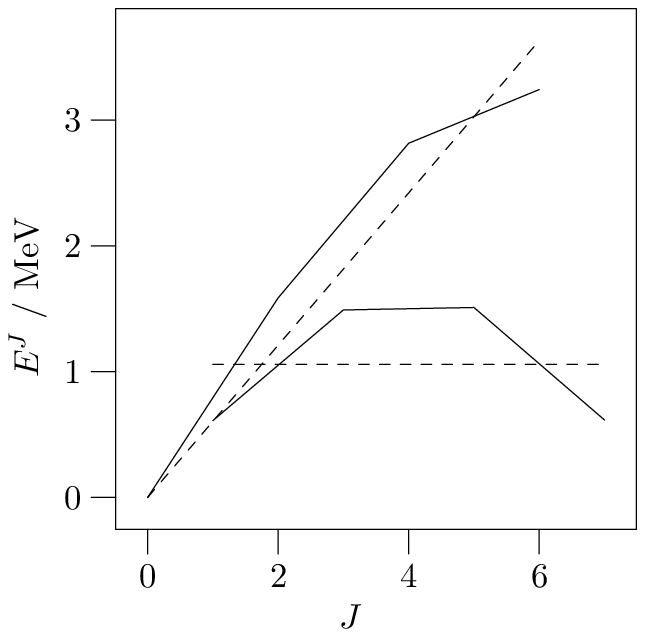}}
  \caption{Empirical interaction of two nucleons in the $1f_{7/2}$
    shell derived from the spectrum of $^{42}$Sc. The matrix elements
    $E^J$ are connected by broken lines separately for even and odd
    J. The dashed lines suggest an approximation of the even $J$
    matrix elements by a function linear in $J$ and the odd $J$ matrix
    elements by a constant.}
\end{figure}

In our present investigation $j$ is arbitrary, and we adopt a
schematic interaction. The nuclei considered are such which have two
protons and two neutrons in the given shell. It is well known that
such a model also applies to the case of two proton holes and two
neutron holes. Our choice of schematic interaction is motivated by the
gross structure of the matrix elements of Ref.~\refcite{ref:ZaRo},
which are displayed in \fref{fig:42Sc}. Shown there are the
interaction matrix elements $E^J = \langle (jj)J | v | (jj)J\rangle$,
where $j = 7/2$. It is seen that while the even $J$ matrix element
rises steeply with $J$, the odd $J$ matrix element varies much less
and its average slope as a function of $J$ is approximately zero. This
suggests to approximate the even $J$ matrix elements by a function
linear in $J$ and the odd $J$ matrix elements by a constant $c$. The
only effect of this constant is to add $(3 - \frac12 T(T+1))c$ to all
energies, where $T$ is the total isospin. The stationary wave
functions are not affected. As we consider only states with $T = 0$,
we can therefore choose $c = 0$ just as well. The interaction then
depends only on an energy scale factor. Choosing this scale factor in
the simplest possible way we arrive at the following schematic
interaction to be studied in the subsequent part of this chapter.
\begin{equation}
  E^J = \begin{cases}
    J , & \text{even $J$,} \\ 0 , & \text{odd $J$.} \end{cases}
\end{equation}

The next section shows examples of results derived numerically from
this interaction. We illustrate, in particular, the occurrence for
certain values of $j$ and the total angular momentum $I$, of
stationary states where the sum $J_p + J_n$ of the angular momenta of
the proton and neutron pairs is conserved. We also illustrate that
these states, which we call \textit{special} states, always have
absolute energies (that is, energies before the ground state energy is
subtracted to give an excitation energy) equal to $3(J_p + J_n)/2$. To
finish the section we report a systematic search of special states for
all $j \le 15/2$ and give empiric rules for their occurrence. In
\sref{sec:expl} we then explain these observations, and the present
chapter is summarised in \sref{sec:sum}.

\section{\label{sec:num}Numeric results}

\begin{figure}\label{fig:yen}
  \centerline{\includegraphics{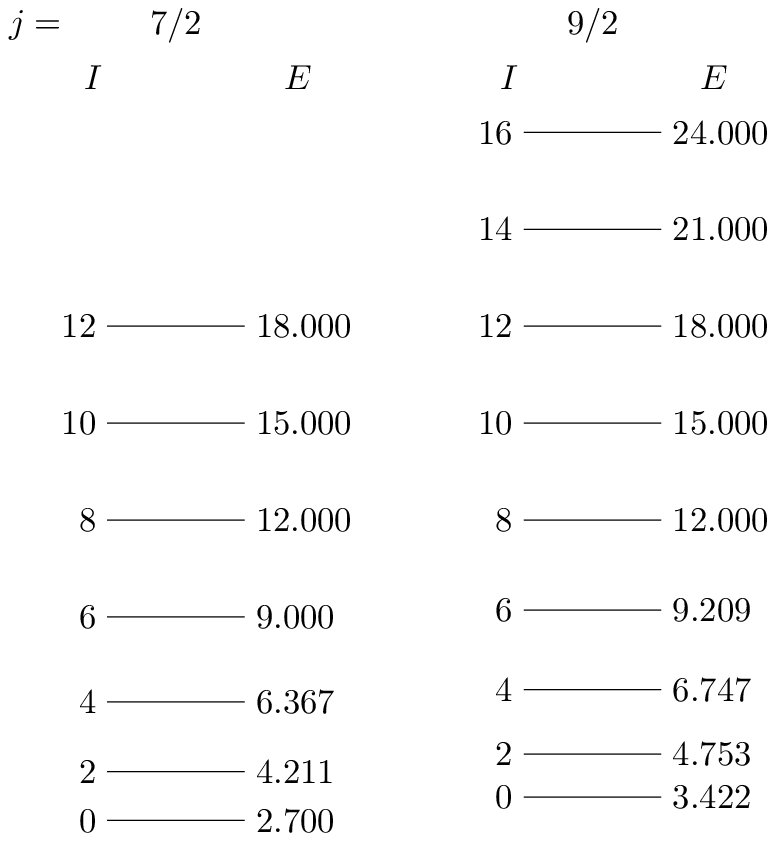}}
  \caption{Calculated even $I$ yrast bands for $j = 7/2$ and 9/2. The
    absolute energy $E$ of each level is indicated.}
\end{figure}

Figure~\ref{fig:yen} shows the even $I$ yrast bands calculated for %
$j = 7/2$ and 9/2. The top half of each band is seen to be strictly
linear. In fact the absolute energies equal $3I/2$. The wave
functions, shown in \tref{tbl:9h_ywf} for $j=9/2$, have a very simple
structure. As all these states have $T = 0$, which implies that the
coefficient of a basic state
\begin{equation}\label{eq:|>}
  | J_pJ_n \rangle = | ((jj)J_p(jj)J_n)IM \rangle ,
\end{equation}
acquires a sign factor $(-)^I$ when $J_p$ and $J_n$ are interchanged,
we show in the table the coefficients of the basic states
\begin{equation}\label{eq:|>e}
  |J_pJ_n\rangle_{\text e} = 2^{-\frac{1+\delta_{J_pJ_n}}2}
    ( |J_pJ_n\rangle + (-)^I |J_nJ_p\rangle ) .
\end{equation}
All the states listed in \tref{tbl:9h_ywf} are seen to have only
components with $J_p + J_n = I$. In \eref{eq:|>} the first two angular
momenta $j$ are those of the individual protons and the last two those
of the neutrons. The total magnetic quantum number $M$ is arbitrary.
In \eref{eq:|>e} the angular momenta $J_p$ and $J_n$ are even, %
$J_p \ge J_n$ for even $I$ and $J_p > J_n$ for odd $I$. The subscript
`e' stands for `even' to indicate that these states span the space
where $T$ is even for the given $j$, $I$ and $M$. This is used in
\sref{sec:expl}.

\begin{table}
  \tbl{\label{tbl:9h_ywf}Wave functions in the calculated even $I$
     yrast band for $j = 9/2$ and $I \ge 8$. Shown are the
     coeeficients of the  states $| J_pJ_n \rangle_{\text e}$ defined
     by \eref{eq:|>e}.}
  {\begin{tabular}{@{}cccccccc@{}}
    \toprule
    $J_p$ & $J_n$ & $I=$ & 8 & 10 & 12 & 14 & 16 \\
    \colrule
    4 & 4 && 0.595 \\
    6 & 2 && 0.700 \\
    6 & 4 && 0.000 & 0.885 \\
    6 & 6 && 0.000 & 0.000 & 0.745 &  &  \\
    8 & 0 && 0.395 \\
    8 & 2 && 0.000 & 0.466 \\
    8 & 4 && 0.000 & 0.000 & 0.667 \\
    8 & 6 && 0.000 & 0.000 & 0.000 & 1.000 \\
    8 & 8 && 0.000 & 0.000 & 0.000 & 0.000 & 1.000 \\
    \botrule
  \end{tabular}}
\end{table}

Several other states are degenerate with these even $I$ yrast states.
They are listed in \tref{tbl:9h_ny}. All these states have $T = 0$. As
this holds for all the states discussed in this chapter, we do not
mention it any more. Most of the states in \tref{tbl:9h_ny} have odd
$I$. The lowest state for each of $I = 9$, 11 and 13 is an yrast state
and degenerate with the yrast state with one unit higher angular
momentum. (The only state with $I = 15$, which as such is necessarily
the yrast state for this angular momentum, has $T = 1$.) Inspecting
the wave functions, one notices again a conservation of $J_p + J_n$.
Furthermore the energy is always $3(J_p + J_n)/2$

\newbox\msgn\setbox\msgn\hbox{-}\newcommand\mns{\hskip-\wd\msgn-}
\begin{table}
  \tbl{\label{tbl:9h_ny}Energies $E$ and wave functions of $j = 9/2$
    special states not belonging to the even $I$ yrast band. The wave
    functions are shown as coeeficients of the states
    $| J_pJ_n \rangle_{\text e}$.}
  {\begin{tabular}{@{}cccccccccc@{}}
    \toprule
    && $E=$ & 15 & 15 & 18 & 18 & 21 & 21 & 24 \\
    $J_p$ & $J_n$ & $I=$ & 7 & 9 & 10 & 11 & 11 & 13 & 14 \\
    \colrule
    6 & 2 && 0.000 \\
    6 & 4 && 0.872 & 0.459 & 0.000 \\
    6 & 6 &&&& 0.689 \\
    8 & 0 && 0.000 \\
    8 & 2 && \mns0.489 & 0.888 & 0.000 \\
    8 & 4 && 0.000 & 0.000 & \mns0.725 & 1.000 & 0.000 \\
    8 & 6 && 0.000 & 0.000 & 0.000 & 0.000 & 1.000 & 1.000 & 0.000 \\
    8 & 8 &&&& 0.000 &&&& 1.000 \\
    \botrule
  \end{tabular}} 
\end{table}

An analogous situation emerges for any $j$ we have examined.
Table~\ref{tbl:search} shows the result of a complete search of
special states for $j \le 15/2$. Always the absolute energy is %
$3(J_p + J_n)/2$. The following systematics is inferred from
\tref{tbl:search}.

\pagebreak

\begin{itemize}
\item[Rule 1:] For a given $j$ there is a special state for any $I$
  from $2j - 1$ to $4j - 2$ except $4j - 3$ (which is impossible for
  $j = 1/2$ and accommodates for $j \ge 3/2$ just a single $T = 1$
  state). These states have $J_p + J_n = I$ for even $I$ and %
  $J_p + J_n = I + 1$ for odd $I$ and are yrast states.
\item[Rule 2:] Besides, there are special states with %
  $(J_p + J_n,I) = (4j - 8,4j - 11)$, $(4j - 6,4j - 8)$, %
  $(4j - 4,4j - 7)$ and $(4j - 2,4j - 4)$ provided this $I$ is not
  negative.
\end{itemize}
These rules have only two exceptions, both of which occur for fairly
low $j$: First, there is no %
$(J_p + J_n,I) = (4j - 6,4j - 8) = (4,2)$ special state
for\linebreak$j=5/2$. Second, there is an additional %
$(J_p + J_n,I) = (10,3) =\linebreak(4j - 4,4j - 11)$ special state
for $j = 7/2$.

\begin{table}
  \tbl{\label{tbl:search}All special states occurring for %
    $j \le 15/2$.}
  {\begin{tabular}{@{}ccc|ccc@{}}
    \toprule
    $j$ & $J_p + J_n$ & $I$ & $j$ & $J_p + J_n$ & $I$ \\
    \colrule
    1/2 & 0 & 0 &13/2 & 12 & 12 \\
    3/2 & 2 & 2 & & 14 & 13, 14 \\
    & 4 & 2, 4 & & 16 & 15, 16 \\
    5/2 & 4 & 4 && 18 & 15, 17, 18 \\
    & 6 & 3, 5, 6 && 20 & 18, 19, 20 \\
    & 8 & 6, 8 & & 22 & 19, 21, 22 \\
    7/2 & 6 & 3, 6 && 24 & 22, 24 \\
    & 8 & 6, 7, 8 &15/2 & 14 & 14 \\
    & 10 & 3, 7, 9, 10 && 16 & 15, 16 \\
    & 12 & 10, 12 && 18 & 17, 18 \\
    9/2 & 8 & 8 & & 20 & 19, 20 \\
    & 10 & 7, 9, 10 && 22 & 19, 21, 22 \\
    & 12 & 10, 11, 12 && 24 & 22, 23, 24 \\
    & 14 & 11, 13, 14 & & 26 & 23, 25, 26 \\
    & 16 & 14, 16 & & 28 & 26, 28 \\
    11/2 & 10 & 10 \\
    & 12 & 11, 12 \\
    & 14 & 11, 13, 14 \\
    & 16 & 14, 15, 16 \\
    & 18 & 15, 17, 18 \\
    & 20 & 18, 20 \\   
    \botrule
  \end{tabular}}
\end{table}

The four degenerate levels with $J_p + J_n = 10$ and $I = 3$, 7, 9 and
10 occurring for $j = 7/2$ are familiar from studies by Robinson and
Zamick\cite{ref:RoZa1,ref:RoZa2}. These authors consider an
interaction in the $1f_{7/2}$ shell with $E^J = 0$ for odd $J$ and
arbitrary $E^J$ is for even $J$. (As noted in the introduction, their
results then apply essentially unaltered to the case when $E^J$ is
constant for odd $J$.) From properties of 9-$j$ symbols they derive in
Ref.~\refcite{ref:RoZa1} that for these $I$ there is a stationary
state whose wave function is just $|64\rangle_{\text e}$. Because for
all these $I$ this is the only $|J_pJ_n\rangle_{\text e}$ with %
$J_p + J_n = 10$, these are the same states as considered presently. A
slight extension of the arguments in Ref.~\refcite{ref:RoZa1} shows
that for the more general interaction considered there they have
energies $3(E^6 + E^4)/2$, so they are degenerate. In
Ref.~\refcite{ref:RoZa2} the properties of 9-$j$ symbols employed in
Ref.~\refcite{ref:RoZa1} are derived from the fact that none of the
four angular momenta accomodate $T = 2$. It is shown in
\sref{sec:expl} that when this happens and $E^J = 0$ for odd $J$, then
quite generally any $|J_pJ_n\rangle_{\text e}$ is a stationary state.
Its energy is $3(E^{J_p} + E^{J_n})/2$.

\section{\label{sec:expl}Explanation}

How is it possible that $J_p + J_n$ is conserved in some stationary
states of our schematic Hamiltonian, and why do these states always
have energy\linebreak$3(J_p + J_n)/2$? In order to see how this comes
about notice that for given $j$, $I$ and $M$ this Hamiltonian $H$ has
matrix elements
\begin{multline}\label{eq:H}
  \langle J_pJ_n | H | J_p'J_n'\rangle
  = \delta_{J_pJ_p'}\delta_{J_nJ_n'} ( E^{J_p} + E^{J_n} ) \\
    + 4 \sum_{J_1J_2} \langle J_pJ_n \Vert J_1J_2 \rangle
      E^{J_1} \langle J_1J_2 \Vert J_p'J_n' \rangle ,
\end{multline}
where $\langle J_1J_2 \Vert J_1'J_2' \rangle$ is shorthand for a
unitary 9-$j$ symbol,
\begin{equation}\label{eq:un9j}
  \langle J_1J_2 \Vert J_1'J_2' \rangle
  = \langle ((j_1j_2)J_1(j_3j_4)J_2)IM 
    | ((j_1j_3)J_1'(j_2j_4)J_2'))IM \rangle ,
\end{equation}
where all $j$'s equal $j$. While the angular momenta $J_p$, $J_n$,
$J_p'$ and $J_n'$ are even, $J_1$ and $J_2$ take all values allowed by
the triangle inequalities. It is convenient to define an operator $X$
such that
\begin{equation}\label{eq:X}
  \langle J_1J_2 | X | J_1'J_2' \rangle
  = \langle J_1J_2 \Vert J_1'J_2' \rangle.
\end{equation}
The space with even $T$ is spanned by the states
$|J_pJ_n\rangle_{\text e}$. By the symmetry of %
$\langle J_pJ_n \Vert J_1J_2 \rangle$ the matrix element $\langle
J_pJ_n|_{\text e}\, X|J_1J_2 \rangle$ vanishes unless $J_1$ and $J_2$
have equal parities. Therefore, in the even $T$ space, when %
$E^{J} = 0$ for odd $J$, only even $J_1$ and $J_2$ contribute to the
sum in~\eqref{eq:H}, and we have
\begin{equation}
  H = \Omega + 2 W \Omega W
\end{equation}
with operators $\Omega$ and $W$ acting within the even $T$ space and
defined by
\begin{gather}
  \langle J_pJ_n | \Omega | J_p'J_n' \rangle_{\text e}
  = \delta_{J_pJ_p'} \delta_{J_nJ_n'} ( E^{J_p} + E^{J_n} ) , \\
  \label{eq:W}\langle J_pJ_n | W | J_p'J_n'\rangle_{\text e}
  = \langle J_pJ_n | X | J_p'J_n' \rangle_{\text e} .
\end{gather}
The subscript `e' indicates that the matrix element is taken between
states $| J_pJ_n \rangle_{\text e}$.

We denote by $(ik)$ the interchange of the states of the $i$th and
$k$th nucleons, where the nucleons are numbered in the order of
appearance of their angular momenta in \eref{eq:|>}. Due to %
$(12)| J_pJ_n \rangle = (34)| J_pJ_n \rangle = - | J_pJ_n \rangle$
one can make in \eref{eq:W} the substitution
\begin{equation}
  4 X = (13) + (14) + (23) + (24) .
\end{equation}
By Eq.~(4) of Ref.~\refcite{ref:Ne} we have
\begin{equation}\label{eq:sumof()}
  \sum_{i < k} (ik) = 4 - 4^2/4 - T(T + 1) = -T(T + 1) .
\end{equation}
As a result the matrix $W$ has the eigenvalue %
$ (- T(T + 1) - 2 \times (-1) )/4 = 1/2$ for $T=0$. In particular, if
some $T = 0$ state is an eigenstate of $\Omega$ it is an eigenstate
of $H$ with eigenvalue $1+2\times(1/2)^2=3/2$ times that of $\Omega$.

This explains the finding of Robinson and Zamick in
Ref.~\refcite{ref:RoZa2}. If $T = 2$ is not accomodated for the given
$j$ and $I$ then the states $| J_pJ_n \rangle_{\text e}$ have $T = 0$.
They are also eigenstates of $\Omega$ with eigenvalue $E^{J_p}
+E^{J_n}$. Therefore they are eigenstates of $H$ with eigenvalue
$3(E^{J_p} +E^{J_n})/2$.

For the Hamiltonian presently considered any linear combination of
states $| J_pJ_n \rangle_{\text e}$ with $J_p + J_n = k$, where $k$ is
a constant, is an eigenstate of $\Omega$ with eigenvalue $k$. What
then remains to be explained is that for the combinations of $j$, $k$
and $I$ obeying the above rules 1 and 2 with the two exceptions
mentioned, there exist such linear combinations which have $T = 0$.
The rest of this section is devoted to a proof of this. The proof is
divided into separate parts for the two rules. Notice that the second
exception is explained already. The special state with %
$(j,k,I) = (7/2,10,3)$ is one of the states discussed by Robinson and
Zamick in Refs.~\refcite{ref:RoZa1,ref:RoZa2}. An explanation of the
first exception is deferred to \sref{sec:rule2}.

\subsection{Rule 1}

We discuss the cases of even and odd $I$ separately.

\paragraph{Even $I$}

Let 
\begin{equation}
  | \psi \rangle = \sum_{J_p + J_n = k} c_{J_p} | (J_pJ_n)kk \rangle,
\end{equation}
with some set of coefficients $c_{J}$, where we have included
explicitly $I$ and $M$ on the left hand side of \eref{eq:|>}. This
state evidently has $I=k$. We assume $k \ge 2j - 1$, so the range
$\mathcal S$ of $J_p$ in the summation is the set of even integers $J$
with $k - 2j + 1 \le J \le 2j - 1$. From formulas for vector coupling
coefficients~\cite{ref:Ed} one gets
\begin{multline}
  \langle m_1m_2m_3m_4 | \psi \rangle \mathrel{\mathop:}=
    ( \langle jm_1 | \times \langle jm_2 | \times
      \langle jm_3 | \times \langle jm_4 | ) \, | \psi \rangle \\
  = \delta_{m_1 + m_2 + m_3 + m_4,k} (-)^{m_1 - m_3}
    a(m_1) a(m_2) a(m_3) a(m_4) f(m_1 + m_2),
\end{multline}
where the $m$'s are single nucleon magnetic quantum numbers, and
\begin{gather}
  a(m) = \sqrt{\frac{(j+m)!}{(j-m)!}} , \\
  f(\mu) = \begin{cases}
    b(\mu) b(k-\mu) c_\mu , & \mu \in \mathcal S ,\\
    0 , & \text{otherwise} ,
  \end{cases}\\
  b(J) = \frac 1{J!} \sqrt{\frac{(2j-J)!(2J+1)!}{(2j+J+1)!}} .
\end{gather}

By \eref{eq:sumof()} the state $|\psi\rangle$ has $T=0$ when it
belongs to the kernel of
\begin{equation}
  K = (13) + (14) + (23) + (24) - 2 .
\end{equation}
This is seen to be equivalent to 
\begin{multline}\label{eq:f-cond}
  (-)^{m_3 - m_1} f(m_3 + m_2) + (-)^{m_4 - m_3} f(m_4 + m_2) \\
  + (-)^{m_1 - m_2} f(m_1 + m_3) + (-)^{m_1 - m_3}f(m_1 + m_4) \\
  - 2 (-)^{m_1 - m_3} f(m_1 + m_2) = 0
\end{multline}
for $m_1 +m_2 + m_3 + m_4 = k$. Equation~\eqref{eq:f-cond} holds when
$f(\mu)$ is constant for $\mu \in \mathcal S$. Indeed, when %
$m_1 + m_2 + m_3 + m_4 = k$, no sum $\mu$ of two of the $m$'s is
greater that $2j$ or less than $k - 2j$, so $\mu \in \mathcal S$ if
$\mu$ is even. First assume that $m_1 + m_2$ is even. If $m_4 + m_2$
is even then the sign factor in the second term in \eref{eq:f-cond}
becomes $(-)^{m_1 - m_3}$. If it is odd, the term vanishes. If %
$m_1 + m_3$ is even, the sign factor in the third term becomes
$(-)^{m_1 - m_3}$. If it is odd, the term vanishes. All sign factors
are thus effectively equal to $(-)^{m_1 - m_3}$. Because with even
$m_1 + m_2$ the sum $m_3 + m_4$ is also even and the $m$'s are
half-integral, the numbers $m_3 + m_2$ and $m_4 + m_2$ have opposite
parities. So do the numbers $m_1 + m_3$ and $m_1 + m_4$. Therefore the
equation hold. If $m_1 + m_2$ is odd, because $m_3 + m_4$ is also odd,
all of $m_3 + m_2$, $m_4 + m_2$, $m_1 + m_3$ and $m_1 + m_4$ have the
same parities. If all of them are even, $m_1 + m_4$, in particular, is
even, so $(-)^{m_3 - m_1} (-)^{m_4 - m_3} = (-)^{m_4 - m_1} = -$.
Similarly, because $m_2 + m_3$ is even, %
$(-)^{m_3 - m_1} (-)^{m_1 - m_2} = (-)^{m_3 - m_2} = -$. So again the
equation holds.

Thus $|\psi\rangle$ is special when 
\begin{equation}\label{eq:c,evenI}
  c_J \propto \frac1{b(J)b(k-J)} .
\end{equation}

\paragraph{Odd $I$}

We now consider a state 
\begin{equation}
  | \psi \rangle = \sum_{J_p + J_n = k}
    c_{J_n} | (J_pJ_n)(k-1)(k-1) \rangle ,
\end{equation}
which has $I = k - 1$, and we assume so far again $k \ge 2j - 1$. This
limit is going to be sharpened. For $k - 1$ to be non-negative
necessarily $k \ge 2$. We also assume
\begin{equation}\label{eq:c-sym}
  c_J = -c_{k-J}
\end{equation}
as required for $T$ to be even. This rules out $k = 4j - 2$ because in
that case $\mathcal S$ has only one element $J = 2j - 1$, whose $c_J$
would then vanish. (It was noted already, indeed, that %
$I = (4j - 2) - 1 = 4j - 3$ acomodates only a single $T = 1$ state.)
Using again formulas from Ref.~\refcite{ref:Ed} we then get
\begin{multline}\label{eq:<m|psi>}
  \langle m_1m_2m_3m_4 | \psi \rangle
  = \delta_{m_1 + m_2 + m_3 + m_4,k - 1} 
    (-)^{m_1-m_3} \sqrt{\frac1{2k}} \\
    a(m_1) a(m_2) a(m_3) a(m_4) g(m_1 + m_2)
    \times \begin{cases}
       m_3 - m_4 , & \text{even $m_1 + m_2$} , \\
       m_1-m_2, & \text{odd $m_1 + m_2$} ,
    \end{cases}
\end{multline}
with 
\begin{gather} \label{eq:g}
  g(\mu) = \begin{cases}
    d(\mu) d(k - \mu) c_\mu , & \mu \in \mathcal S , \\
    g(k - 1 - \mu) , & k - 1 - \mu \in \mathcal S , \\
    0 , & \text{otherwise} ,
  \end{cases} \\
  d(J) = \sqrt J \, b(J).
\end{gather}

As $\langle m_1m_2m_3m_4 | \psi \rangle$ vanishes unless %
$m_1 + m_2 + m_3 + m_4 = k - 1$, this is understood in the following.
The state $|\psi \rangle$ is even under the permutation $(13)(24)$ and
odd under $(12)$, both of which commute with $K$. (That %
$|\psi \rangle$ is even under $(13)(24)$ is seen explicitly from
Eqs.~\eqref{eq:<m|psi>} and \eqref{eq:g}. Quite generally a state with
definite $T$ of equally many protons and neutrons has the parity
$(-)^T$ under the exchange of the entire states of the proton and
neutron subsystems.) Because the $m$'s are half-integral, we can
therefore assume without loss of generality that $m_1 + m_2$ and %
$m_1 + m_3$ are even. Then $m_3 + m_4$, $m_2 + m_4$ and $m_2+m_3$ are
odd and $m_1+m_4$ is even. A sufficient condition for $| \psi \rangle$
to belong to the kernel of $K$ is then
\begin{multline}
  (-)^{m_3 - m_1} (m_3 - m_2) g(m_3 + m_2)
  + (-)^{m_4 - m_3} (m_4-m_2) g(m_4 + m_2) \\
  + (-)^{m_1 - m_2} (m_2 - m_4) g(m_1 + m_3)
  + (-)^{m_1 - m_3} (m_3 - m_2) g(m_1 + m_4) \\
  - 2 (-)^{m_1 - m_3} (m_3 - m_4) g(m_1 + m_2) = 0 .
\end{multline}
By $(-)^{m_3 - m_1} (-)^{m_4 - m_3} = (-)^{m_4 - m_1} = -$, %
$(-)^{m_3 - m_1} (-)^{m_1 - m_2} = (-)^{m_3 - m_2} = +$ and %
$g(\mu) = g(k - 1 - \mu)$ this is reduced to
\begin{equation}\label{eq:g-cond}
  (m_3 - m_2) g(m_3 + m_2) + (m_2 - m_4) g(m_2 + m_4)
  + (m_4 - m_3) g(m_4 + m_3) = 0 .
\end{equation}
An odd sum $\mu$ of two $m$'s cannot be greater than $2j$ or less
than\linebreak$k - 1 - (2j - 1) = k - 2j$. If $\mu = 2j$ both $m$'s
equal $j$, which eliminates the term with this $g(\mu)$ from
\eref{eq:g-cond}. For $k - 2j \le \mu \le 2j - 2$ the number %
$k - 1 - \mu$ belongs to $\mathcal S$. Equation~\eqref{eq:g-cond}
holds if $g(\mu)$ is a polynomial of first degree in $\mu$ for odd
sums $\mu$ of two $m$'s, and it is by the preceding remark sufficient
that $k - 1 - \mu \in \mathcal S$. This is by %
$g(\mu) = g(k - 1 - \mu)$ equivalent to $g(\mu)$ being a polynomial of
first degree in $\mu$ for $\mu\in\mathcal S$.

Choosing 
\begin{equation}\label{eq:c,oddI}
  c_J \propto \frac{k - 2J}{d(J)d(k-J)},
\end{equation}
gives the polynomial $g(\mu) \propto k - 2\mu$ of first degree, which
satisfies \eref{eq:c-sym}. The state $| \psi \rangle$ is then special.
As the denominator in \eref{eq:c,oddI} vanishes for $J = 0$, this must
not be allowed. Then $k = 2j - 1$ is ruled out and the final scope of
the proof is $2j + 1 \le k \le 4j - 4$, corresponding to odd $I$ with
$2j \le I \le 4j - 5$.

\subsection{\label{sec:rule2}Rule 2}

We introduced already the notion of the \textit{even $T$ space}, which
is the space of states with given $j$, $I$ and $M$ and even $T$. The
condition $J_p + J_n = k$ defines a subspace, which we call the
\textit{$k$ space}. Its dimension is called the %
\textit{$k$ dimension}. The condition $T = 0$ similarly defines a
subspace. This we call the \textit{$T = 0$ space} and its dimension
the \textit{$T = 0$ dimension}. A \textit{$T = 2$ space} and a
\textit{$T = 2$ dimension} are defined analogously. If the $k$
dimension is greater than the $T = 2$ dimension then at least one
state in the $k$ space is perpendicular to the $T = 2$ space and thus
belongs to the $T = 0$ space. It is then a special state. A special
state thus exist for given $j$, $k$ and $I$ whenever the $k$ dimension
exceeds the $T = 2$ dimension. Note that this is a sufficient but not
a necessary condition. As we shall see, is not satified in some cases
covered by rule 1.

In particular, if the $T = 2$ space is zerodimensional then each
entire $k$ space consists of special states. It turns out, as
discussed below, that the $k$ dimension never exceeds the $T = 2$
dimension by more that one, so in that case any positive $k$ dimension
is just one. That is, the $k$ space is spanned by a single %
$| J_pJ_n \rangle_{\text e}$. These are the states discussed by
Robinson and Zamick in Refs.~\refcite{ref:RoZa1,ref:RoZa2}

The $k$ and $T = 2$ dimensions are determined by combinatorics. In
particular, because a state $| \psi \rangle$ has $T = 2$ if and only
if $\langle m_1m_2m_3m_4 | \psi \rangle$ is antisymmetric in the
$m$'s, the $T = 2$ dimension is given as the number of combinations of
$m_1 > m_2 > m_3 > m_4$ such that $\sum m = I$. The counts are
simplified if one assumes $I \ge 2j - 1$ because then, in counting the
combinations of $J_p \ge J_n$ that give $J_p + J_n = k$ and the
combinations of $m_1 > m_2 > m_3 > m_4$ that give %
$\sum m = I$, one can neglect the lower limits $J_n \ge 0$ and %
$m_4 \ge -j$. The condition $I \ge 2j - 1$ also secures the triangle
inequality $J_p \le J_n + I$. The triangle inequality %
$I \le J_n + J_p$ is secured by $k \ge I$. Therefore, %
if $I \ge 2j - 1$ the $k$ dimension is a function of $x = 4j - k$ and
the $T = 2$ dimension a function of $y = 4j - I$.

The following tables show the result of this combinatoric analysis.

{\centering

\medskip

\begin{tabular}{ccccccccc}
  $x$ & 2 & 4 & 6 & 8 & 10 & 12 & 14 & 16 \\
  \hline 
  $k$ dim., even $y$ & 1 & 1 & 2 & 2 & 3 & 3 & 4 & 4 \\
  $k$ dim., odd $y$ & 0 & 1 & 1 & 2 & 2 & 3 & 3 & 4
\end{tabular}

\medskip

\begin{tabular}{ccccccccccccc}
  $y$ & 6 & 7 & 8 & 9 & 10 & 11 & 12 & 13 & 14 & 15 & 16 & 17 \\
  \hline 
  $T = 2$ dim. & 1 & 0 & 1 & 1 & 2 & 1 & 3 & 2 & 4 & 3 & 5 & 4 \\
\end{tabular}

\medskip

}

\noindent Because $J_p,J_n \le 2j - 1$ both $x$ and $y$ are at least
2. The $T = 2$ dimension vanishes for $y < 6$ because no combination
of four different $m$'s have a sum greater than $4j - 6$. The
condition $k \ge I$ translates to $x \le y$. It is evident that the %
$T = 2$ dimension rises more rapidly than the $k$ dimension with
increasing $y$ so that the values of $x$ and $y$ included in the
tables suffice to determine the cases when the latter dimension
exceeds the former.

This is seen to happen when $y = x \le 10$ or %
$4 \le x = y - 1 \le 12$, which corresponds to rule 1 with an
additional upper limit on $x$. As rule 1 does not have this upper
limit, it is thus more general than can be inferred from this
dimensional analysis. The only other cases when the $k$ dimension
exceeds the $T = 2$ dimension are $(x,y) = (2,4)$, $(4,7)$, $(6,8)$
and $(8,11)$, which correspond exactly to rule 2.

It was assumed that $I \ge 2j -1$, and all the cases of the $k$
dimension exceeding the $T = 2$ dimension that were identified have %
$y \le 13$. When $y \le 13$ the condition $I \ge 2j -1$ is satified
for $j \ge 13/2$. The dimensional analysis is thus exhaustive for
these $j$. The combinations of $j$, $k$ and $I$ that occur for %
$j \le 11/2$ are finite in number, so they and can be examined
individually. This was done in the search of special states with %
$j \le 15/2$ reported in \sref{sec:num}. It turns out that all the
special states with $j \le 11/2$ appear when the $k$ dimension is
greater than the $T = 2$ dimension with one exception: For $j = 11/2$
and $I = k = 10$ both dimensions equal 3. This state is covered by
rule 1, and the equality of the two dimensions is, in fact, consistent
with the combinatoric analysis, which is valid for $j \ge 9/2$ when
$I$ is even and then requires $I \ge 4j - 10$ for the $k$ dimension to
exceed the $T = 2$ dimension. For $(j,I) = (5/2,2)$ both the $T = 2$
dimension and the maximal $k$ dimension equal 1, and the $k = 4$
special state anticipated by rule 2 indeed does not appear.

The tables above show that for $j \ge 13/2$ the $k$ dimension never
exceeds the $T = 2$ dimension by more than one. This is found to hold
also for\linebreak$j \le 11/2$. It can be turned around to say that
the dimension of the configuration space of four identical fermions
with given $j$, $I$ and $M$ is never less than the maximal $k$
dimension minus one. As an empiric rule, a special state is always
unique to the given $j$, $k$, $I$ and $M$, that is, any $k$ space has
at most a onedimensional intersection with the $T = 0$ space.

\subsection{Wave functions}

The wave functions of special states occurring by rule 1 are given by
Eqs.~\eqref{eq:c,evenI} and \eqref{eq:c,oddI}. For the special states
occuring by rule 2 with its two exceptions, the $k$ dimension never
exceeds two. This follows for $I \ge 2j - 1$ from the first table in
\sref{sec:rule2}, and it holds, as well, in the four cases (see
\tref{tbl:search}) with $I < 2j - 1$. If the $k$ dimension is one, the
special state $| \psi \rangle$ is a single %
$| J_pJ_n \rangle_{\text e}$. If the $k$ dimension is two, %
$| \psi \rangle$ is a linear combination
\begin{equation}\label{eq:kdim2}
  | \psi \rangle = \alpha | a \rangle + \beta | b \rangle ,
\end{equation}
where $| a \rangle$ and $| b \rangle$ are states
$| J_pJ_n \rangle_{\text e}$. The ratio of the coefficients $\alpha$
and $\beta$ is determined by the fact that, having $T = 0$, the state
$| \psi \rangle$ is an eigenstate with eigenvalue $1/2$ of the
operator $W$ defined by \eref{eq:W}. Explicitly
\begin{equation}\label{al-be-rat}
  \frac\alpha\beta
    = \frac{\langle a|W|b \rangle}{\frac12 - \langle a|W|a \rangle}
    = \frac{\frac12 - \langle b|W|b \rangle}{\langle a|W|b \rangle} .
\end{equation}
By Eqs.~\eqref{eq:W}, \eqref{eq:X} and \eqref{eq:un9j} the matrix
elements of $W$ are\cite{ref:Ed} 9-$j$ symbols multiplied by factors
$\sqrt{2q +1}$ for some angular momenta $q$ and possibly factors %
$\sqrt 2$.

For any special state occuring by either rule 1 or rule 2 with its two
exceptions, if the $k$ dimension is one the $T = 2$ dimension
vanishes. For $I \ge 2j - 1$ this can be inferred again from the
tables in \sref{sec:rule2}, and again it holds, as well, in the four
cases with $I < 2j - 1$. It implies that for these $j$ and $I$ the
entire even $T$ space has $T = 0$ so that within this space $W$ is
equal to the constant $1/2$ and every $| J_pJ_n \rangle_{\text e}$ is
special. As shown in Ref.~\refcite{ref:RoZa2}, that $W$ is equal to
the constant $1/2$ within the even $T$ space can be inferred also
directly from the fact that these $j$ and $I$ do not accomodate %
$T = 2$. It implies that the 9-$j$ symbols in the matrix elements of
$W$ between different $| J_pJ_n \rangle_{\text e}$ vanish.

When a special state belongs to a twodimensional $k$ space, the
requirement that both expressions in \eref{al-be-rat} give the same
result entails relations between the 9-$j$ symbols involved. So does
the requirement that the matrix element of $W$ between the
state~\eqref{eq:kdim2} and any $| J_pJ_n \rangle_{\text e}$ with a
different $k$ vanishes. The expressions ~\eqref{eq:c,evenI} and
\eqref{eq:c,oddI} give rise to similar relations involving several
9-$j$ symbols.

\section{Summary}\label{sec:sum}

We studied the system of two protons and two neutrons in a $j$ shell
with the two nucleon interaction matrix element equal to the two
nucleon angular momentum $J$ for even $J$ and zero for odd $J$. This
model has a straightforward generalisation to the case when the matrix
element is linear in $J$ for even $J$ and constant for odd $J$. It was
found to exhibit for any $j \ge 3/2$ several stationary states where
the sum $J_p + J_n$ of the angular momenta of the proton and neutron
pairs is conserved. The absolute energies of these states, which we
call \textit{special}, that is, their energies before the ground state
energy is subtracted to give excitation energies, are %
$3(J_p + J_n)/2$. Special states in particular form the even and odd
$I$ yrast bands from $I = 2j -1$ to the maximal $I = 4j - 2$ except %
$I = 4j - 3$, where $I$ is the total angular momentum. Other,
non-yrast states are also special.

It was shown that any state which conserves $J_p + J_n$ is in this
model a stationary state with absolute energy $3(J_p + J_n)/2$
provided it has isospin $T = 0$. Using explicit expressions for vector
coupling coefficients we then demonstrated that such states exist for
all the yrast total angular momenta $I$ specified above. The non-yrast
special states could be explained by a combinatoric analysis of the
dimensions of various subspaces of the configuration space. Explicit
expressions for the wave functions of all special states are provided
by our study.

\section*{Acknowledgments}

Wesley Pereira is a student at Essex College, Newark, New Jersey,
07102. His research at Rutgers is funded by a Garden State Stokes
Alliance for Minorities Participation (G.S.L.S.A.M.P.) internship.

Ricardo Garcia has two institutional affiliations: Rutgers University,
and the University of Puerto Rico, Rio Piedras Campus. The permament
address associated with the UPR-RP is University of Puerto Rico, San
Juan, Puerto Rico 00931. He acknowledges that to carry out this work
he has received support via the Research Undergraduate Experience
program (REU) from the U.S. National Science Foundation through grant
PHY-1263280 and thanks the REU Physics program at Rutgers University
for their support.

\bibliographystyle{ws-rv-van}\bibliography{gerry}

\end{document}